\documentclass[floatfix,aps,amsmath,nofootinbib,twocolumn,10pt]{revtex4}

\usepackage{listings}
\usepackage{graphicx}
\usepackage{bm}
\usepackage{rotating}
\usepackage{array}
\usepackage{amsmath}
\usepackage{amssymb} 
\usepackage{mathrsfs} 
\usepackage{cancel}
\usepackage{color}
\usepackage{appendix}

\def\({\left(}
\def\){\right)}
\def\[{\left[}
\def\]{\right]}

\def\e{\begin{equation}}
\def\q{\end{equation}}
\def\m{\begin{eqnarray}}
\def\n{\end{eqnarray}}
\newcommand{\Mov}[1]{{\color{black}{#1}}}
\newcommand{\Ke}[1]{{\color{black}{#1}}}
\begin{document}

\title{Fuzzy dark matter soliton as gravitational lens}

\author{Ke Wang$^{1}$}
\thanks{{wangke@lnnu.edu.cn}}
\author{M. Le Delliou$^{2,3,4,5,6}$}
\thanks{Corresponding author: {delliou@lzu.edu.cn,\\\phantom{Corresponding author: $l$}Morgan.LeDelliou.IFT@gmail.com}}
\affiliation{$^1$Department of Physics, Liaoning Normal University, Dalian 116029, China}
\affiliation{$^2$Institute of Theoretical Physics $\&$ Research Center of Gravitation, Lanzhou University, Lanzhou 730000, China}
\affiliation{$^3$Key Laboratory of Quantum Theory and Applications of MoE $\&$ Gansu Provincial Research Center for Basic Disciplines of Quantum Physics, Lanzhou University, Lanzhou 730000, China}
\affiliation{$^4$Lanzhou Center for Theoretical Physics $\&$ Key Laboratory of Theoretical Physics of Gansu Province, Lanzhou University, Lanzhou 730000, China}
\affiliation{$^5$Instituto de Astrof\'isica e Ci\^encias do Espa\c co, Universidade de Lisboa,
Faculdade de Ci\^encias, Ed.~C8, Campo Grande, 1769-016 Lisboa, Portugal}
\affiliation{$^6$Universit\'e de Paris-Cit\'e, APC-Astroparticule et Cosmologie (UMR-CNRS 7164), 
F-75006 Paris, France}

\date{\today}

\begin{abstract}
The Schrödinger-Poisson (SP) equations predict fuzzy dark matter (FDM) solitons. Given the FDM mass $\sim10^{-20}\rm~{eV}/c^2$, the FDM soliton in the Milky Way is massive $\sim 10^7~M_{\odot}$ but diffuse $\sim 10{\rm~pc}$. Therefore, such FDM soliton can serve as a gravitational lens for gravitational waves (GWs) with frequency $\sim10^{-8}{\rm~Hz}$. In this paper, we investigate its gravitational lensing effects by numerical simulation of the propagation of GWs through it. We find that the maximum magnification factor of GWs is very small $\sim10^{-4}$, but the corresponding magnification zone is huge $\sim6{\rm~pc}$ for FDM with mass equal to $8\times10^{-21}\rm~{eV}/c^2$. Consequently, this small magnification factor compounding  over such large magnification zone results in a small antisotropy of $\sim10^{-4}$ over a large solid angle in the GW background. That level of antisotropy is out of the sensitivity, $<20\%$, of the pulsar timing arrays today.

\end{abstract}

\pacs{???}

\maketitle


\section{Introduction}
\label{sec:intro}
Rotation curves of galaxies~\cite{Rubin:1982kyu}, evolution of large-scale structure~\cite{Davis:1985rj} and gravitational lensing observations~\cite{Clowe:2006eq} \Mov{favour }
dark matter (DM) \Mov{over }
gravity modifications~\cite{Sanders:2006sz}.
According to the standard Lambda cold DM ($\Lambda$CDM) cosmological model and the latest cosmic microwave background (CMB) observations~\cite{Planck:2018vyg}, DM is cold and accounts for about $26\%$ of today's energy density in the Universe.
However, in fact, CDM does not have an obvious \Mov{observational preference }
over \Mov{the }
warm and hot \Mov{competing DM models}
.
\Mov{This is because }
both 
the weakly interacting massive particles (WIMPs) grounded on supersymmetric theories of particle physics and primordial black holes (BHs) have not yet been detected~\cite{PandaX-II:2016vec,LUX:2015abn,LUX:2016ggv} or identified~\cite{Carr:2016drx}\Mov{. The }
former 
\Mov{remains }
one of the most promising particle candidates while the latter is one of the most promising primordial object candidates for CDM.
In addition to these null \Mov{detection }results, 
some failures of CDM particles on sub-galactic scales~\cite{Primack:2009jr,Bull:2015stt}\Mov{ increase the doubt on CDM. It is therefore }
reasonable to pay attention to some \Mov{of its }alternatives
.


One of \Mov{those }
promising alternatives 
is the ultralight scalar field with spin-$0$, extraordinarily light mass ($\sim10^{-22}\rm~{eV}/c^2$) and de Broglie wavelength comparable to a few kpc, coined fuzzy DM (FDM)~\cite{Hu:2000ke}. It can not only behave as CDM on large scales, but also avoid CDM's small-scale crises~\cite{Hu:2000ke}. Due to its wave nature, FDM can change the pulse arrival time of the pulsar and be detected by pulsar timing arrays~\cite[PTA]{Khmelnitsky:2013lxt}\Mov{ of highly stable millisecond pulsars}.
Besides 
\Mov{tracking }by PTA, many other FDM detection methods are proposed. Similarly to gravitational wave (GW) detection, for example, the direct detection of FDM (or its wind) by space-based laser interferometers such as the Laser Interferometer Space Antenna (LISA)~\cite{LISA:2017pwj} has been estimated~\cite{Aoki:2016kwl,Yu:2023iog}. LISA can also detect FDM indirectly by the frequency modulation of GWs due to FDM~\cite{Wang:2023phr} or other effects, as \Mov{listed }
in a review of GW probes of particle DM~\cite{Miller:2025yyx}. Moreover, FDM can affect orbital motions of astrophysical objects in the galaxy~\cite{Blas:2016ddr,Boskovic:2018rub} and lead to black hole superradiant instability~\cite{Brito:2020lup}.

\Ke{Although a lower bound of order $10^{-19}\rm~{eV}/c^2$ has been proposed~\cite{Amin:2022nlh,Dalal:2022rmp}, obtained from dynamical heating of stars in Ultra Faint Dwarf galaxies~\cite{Dalal:2022rmp} or from the non-detection of power spectrum suppression due to free streaming and white-noise enhancement of density perturbations~\cite{Amin:2022nlh}
}, in this paper, we propose another 
\Ke{radically different} detection method for FDM with mass $\sim10^{-20}\rm~{eV}/c^2$\Ke{, which could potentially provide an independent constraint on the FDM mass}. Due to its large occupation numbers in galactic halos, FDM behaves as a classical field obeying the coupled Schrödinger–Poisson (SP) system of equations Eq.~\eqref{eq:sp}. According to it, FDM can condensate into a ground state of many particles called an FDM soliton.\Ke{\footnote{\Ke{Note that soliton formation is a generic prediction for light bosonic dark matter particles \cite{Jain:2021pnk,Schiappacasse:2025mao} if $m \ll \rm~{eV}/c^2$~\cite{Hui:2021tkt}.}}} Although the FDM soliton is massive $\sim 10^7~M_{\odot}$, it is diffuse $\sim 10{\rm~pc}$. Therefore, \Mov{it opens the possibility for }the FDM soliton \Mov{to }
impose 
some gravitational lensing effects on GWs with frequency $\sim10^{-8}{\rm~Hz}$ propagating through it. In turn, future detection of such signatures on 
lensed GWs \Mov{imprinted }
by the FDM soliton would constrain 
the 
\Mov{FDM particle }propert\Mov{ies}
.
\Ke{For example, the sensitivity band ``sweet spot" for PTA experiments such as NANOGrav~\cite{NANOGrav:2023gor} and the European Pulsar Timing Array~\cite{EPTA:2023fyk} lies around $\sim10^{-8}{\rm~Hz}$, which correspond to the ultra-low-frequency band, a unique window into GWs with wavelength $\sim 1{\rm~pc}$ from the growth of the universe's most massive objects and early cosmological processes. Such critical wavelength, or frequency, is comparable with soliton sizes corresponding to FDM masses around $10^{-20}~\rm{eV/c^2}$, as shown in Fig.~\ref{fig:soliton}.}

\Mov{The mass distribution of }CDM halo\Mov{es, whether following }
a Navarro-Frenk-White (NFW) profile~\cite{Navarro:1995iw} or \Mov{any} other simplified \Mov{density }profiles\Mov{ also presents the possibility of diffuse gravitational lensing. The lensing effect of such CDM distribution }
on GWs has been studied by calculating the corresponding amplification factor and time delay
~\cite{Choi:2021bkx,Gao:2021sxw,Cremonese:2021ahz,Guo:2022dre,Tambalo:2022wlm}.
However, the usual analytic expressions of the amplification factor and time delay~\cite{Takahashi:2003ix} 
depend on the propagation equation for GWs.
In other words, the 
usual expressions are not suitable for the \Mov{GW }propagation equation 
inside the lens, which \Mov{offers a different propagation medium than }
its counterpart outside the lens, as shown in Eq.~\eqref{eq:gw}.
\Mov{For any }
propagation equation for GWs, \Mov{their }direct numerical integration \Mov{
eventually represents }
the correct treatments~\cite{He:2021hhl,Qiu:2022dya,He:2022sjf,Yin:2023kzr}.
\Mov{In the present }
paper, we will numerically integrate the propagation equation for GWs both inside and outside the FDM soliton with a modified $\mathtt{GWsim}$~\cite{He:2021hhl} code, which is further based on the publicly available finite element package $\mathtt{deal.ii}$~\cite{dealii}.  

This paper is organized as follows. In Section~\ref{sec:soliton}, we calculate the soliton profiles for a set of FDM masses. \Mov{We then transform, in }
Section~\ref{sec:lens}, 
the propagation equation for GWs into matrix form. \Mov{Four simulations, in }
Section~\ref{sec:sim}, \Mov{are subsequently performed }
for different FDM soliton lens. Finally, a brief summary and discussion are provided in Section~\ref{sec:sum}.

\section{FDM soliton}
\label{sec:soliton}
FDM obeys the following SP system\Mov{\footnote{\Mov{Our calculations in this section are reproducing the structures of similar calculations in \cite{Guzman:2004wj,Davies:2019wgi,Tan:2024dne}, with modifications to accommodate our particular problems.}}},
\begin{equation}
\label{eq:sp} 
\begin{cases}
\begin{aligned}
&i\hbar\frac{\partial \psi}{\partial t}=\left(-\frac{\hbar^2}{2m}\nabla^2+mc^2\Phi\right)\psi , \\
&c^2\nabla^2 \Phi=4\pi G |\psi|^2, 
\end{aligned}
\end{cases}
\end{equation}
where $m$ is the mass of the FDM particle, $\psi$ 
\Ke{retains the mathematical form of a }
wavefunction\Ke{, whose square norm represents the density 
distribution of 
FDM
}, and $\Phi$ is the dimensionless gravitational potential, 
sourced by the FDM density $\rho=|\psi|^2$. \Mov{Restricting }
to a spherically symmetric system, the waveform features an ansatz \Mov{solution }
$\psi(r, t) =e^{-i\gamma t/\hbar}\phi(r)$, where $\gamma$ is the ansatz energy eigenvalue.
The 
FDM soliton density
$\rho(r)=|\psi|^2=\phi^2(r)$ is \Mov{then }simply related to the FDM soliton mass $M=\int_0^\infty4\pi r^2\rho(r)dr$. 
Meanwhile, with\Mov{in} the spherical ansatz, Eq.~(\ref{eq:sp}) is simplified as
\begin{equation}
\label{eq:sp11} 
\begin{cases}
\begin{aligned}
&\frac{\partial^2 (\tilde{r}\tilde{\phi})}{\partial \tilde{r}^2}=2\tilde{r}\left(\tilde{\Phi}-\tilde{\gamma}\right)\tilde{\phi},\\
&\frac{\partial^2 (\tilde{r}\tilde{\Phi})}{\partial \tilde{r}^2}=\tilde{r}\tilde{\phi}^2,
\end{aligned}
\end{cases}
\end{equation}
\Mov{using the following }
set of dimensionless variables
\begin{align}
&\tilde{\phi}\equiv\frac{\hbar\sqrt{4 \pi G}}{mc^2} \phi,\label{eq:dimLessPsi}\\
&\tilde{r}\equiv\frac{mc}{\hbar}r,\\
&\tilde{\Phi}\equiv\Phi,\\
&\tilde{\gamma}\equiv\frac{1}{mc^2}\gamma,\\
&\tilde{M} \equiv \frac{GMm}{\hbar c}\label{eq:dimLessM}.
\end{align}

\begin{figure}[]
\begin{center}
\includegraphics[width= 9cm]{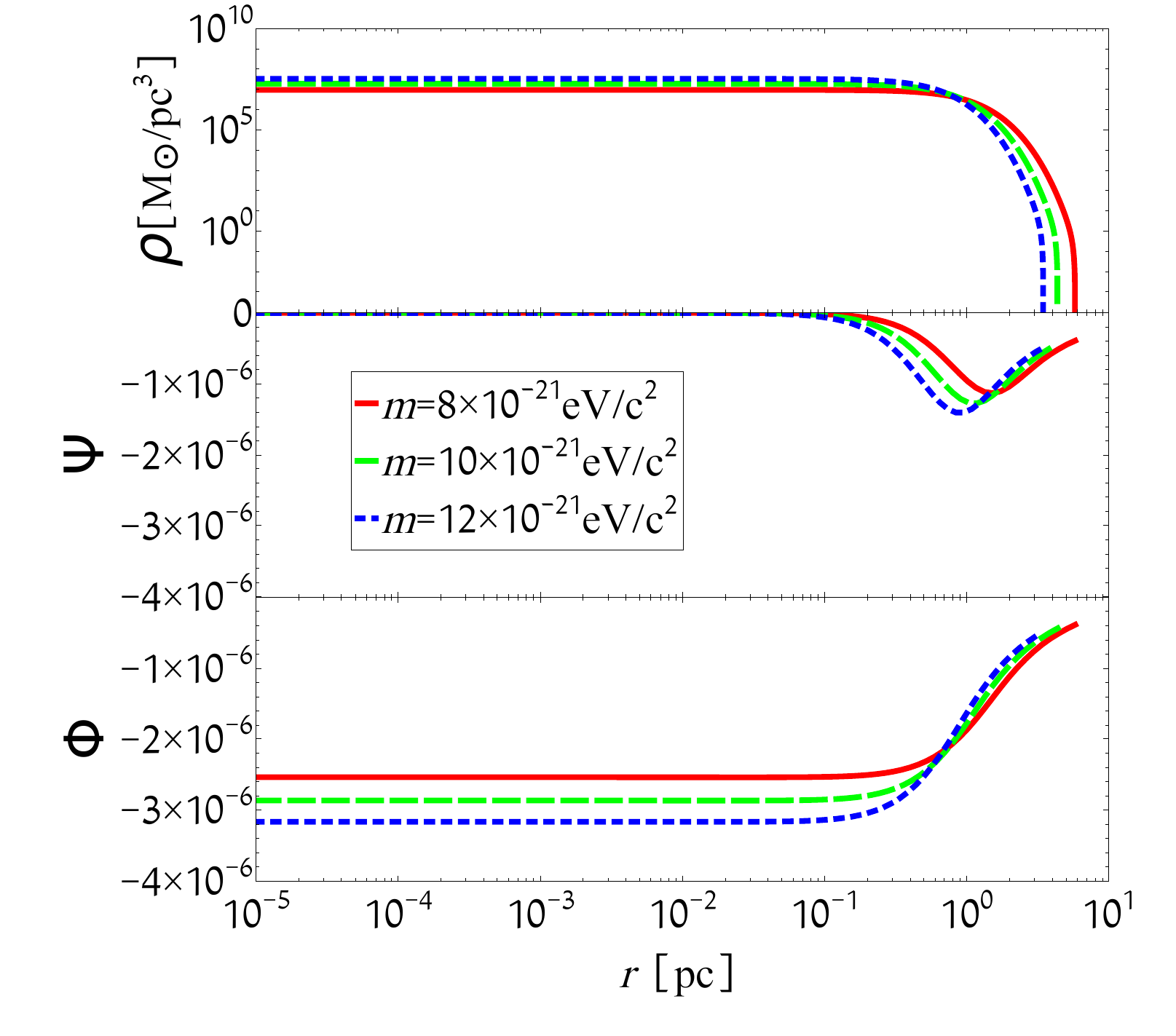}
\end{center}
\caption{FDM soliton profiles for a set of FDM masses $m$, where the gravitational potential $\Phi({\bf{r}})$ decays approximately $\propto-\frac{GM(\bf{r})}{c^2|\bf{r}|}$ and the perturbation of the spatial curvature $\Psi({\bf{r}})=-\frac{GM(\bf{r})}{c^2|\bf{r}|}$\Mov{~\cite{Guzman:2004wj}}.}
\label{fig:soliton}
\end{figure}

\Mov{Selecting }
\begin{enumerate}
    \item the arbitrary normalization $\tilde{\phi}(\tilde{r}=0)=1$,
    \item the boundary conditions $\tilde{\phi}( \tilde{r} = \infty) = 0$,
$\frac{\partial \tilde{\phi}}{\partial \tilde{r}}|_{\tilde{r} = 0} = 0$, $\tilde{\Phi}( \tilde{r} = \infty) = 0$ and $\frac{\partial \tilde{\Phi}}{\partial \tilde{r}}|_{\tilde{r} = 0} = 0$
\end{enumerate}
   and adjusting the quantized eigenvalue $\tilde{\gamma}$, we can calculate the equilibrium configurations from Eq.~(\ref{eq:sp11}) by the shooting method. The only stable solution from the smallest \Mov{value }$\tilde{\gamma}=-0.69223$ is the ground state with mass $\tilde{M}=2.0622$. 
\Mov{Normalised solutions are }
related to their physical counterparts by the following scaling symmetry 
\begin{align}
&\tilde{\phi}  \longrightarrow  \lambda \tilde{\phi}, \\
&\tilde{r}  \longrightarrow  \lambda^{-1/2} \tilde{r},\\
&\tilde{\Phi}  \longrightarrow  \lambda \tilde{\Phi}, \\
&\tilde{\gamma}  \longrightarrow  \lambda \tilde{\gamma}, \\
&\tilde{M} \longrightarrow  \lambda^{1/2} \tilde{M},
\end{align}
where $\lambda$ can be derived from \Mov{the normalisation constraint }$2.0622\lambda^{1/2}\hbar c/Gm=M$\Mov{, for a fixed FDM particle mass $m$ and} given a physical soliton mass $M$.
The soliton mass $M$ can be predicted from the halo mass $M_{\rm{halo}}$ according to the soliton-halo mass relation, whether from the \Mov{result }
of Ref.~\cite{Schive:2014hza}
\begin{equation} 
\label{eq:r_h}
M \approx 1.25\times 10^{9}\left(\frac{M_{\rm{halo}}}{10^{12}M_{\odot}}\right)^{1/3} \left(\frac{m}{10^{-22}{\rm{eV}}/c^2}\right)^{-1}M_{\odot},
\end{equation}
or following the version of Ref.~\cite{Chan:2021bja}
\begin{equation} 
\label{eq:r_h_n}
\begin{aligned}
M \approx &\beta \left(\frac{m}{8\times10^{-23}\rm{eV}/c^2}\right)^{-3/2}\\
&+\left (\frac{M_{\rm{halo}}}{\gamma}\right )^{\alpha}\left(\frac{m}{8\times10^{-23}\rm{eV}/c^2}\right)^{3(\alpha-1)/2}M_{\odot},
\end{aligned}
\end{equation}%
where $\beta=8.00\times 10^{6}~M_{\odot}$, $\gamma=10^{-5.73}~M_{\odot}$ and $\alpha=0.515$. The following discussion \Mov{chooses the latter, as Eq.~\eqref{eq:r_h_n} is the newest version, and }takes 
the Milky Way value\Mov{, }
$M_{\rm{halo}}=1\times 10^{12} ~M_{\odot}$~\cite{Wang:2019ubx}\Mov{, } as an example. \Mov{Accordingly}
, the FDM soliton density profiles for a set of FDM masses are shown in Fig.~\ref{fig:soliton}\Mov{, which is consistent with the dimensionless Fig.~1 of \cite{Guzman:2004wj}}.

\section{FDM soliton as gravitational lens}
\label{sec:lens}
We consider GWs with frequency $f\sim10^{-8}{\rm~Hz}$ propagating through an FDM soliton with size $\sim10{\rm~pc}$. Although the 
\Mov{underlying metric }is coherently oscillating \Mov{because of FDM }with $\omega=2mc^2/\hbar=3\times10^{-8}(mc^2/10^{-23}\rm~eV){\rm~Hz}$, GWs cannot feel the oscillations when $f\ll\omega$. \Mov{We approximate the underlying metric with a static model, }
given by
\begin{equation}
ds^2=-(1+2\Phi)c^2dt^2+(1-2\Psi)d{\bf{r}} ^2\equiv g^{(\rm B)}_{\mu\nu}dx^{\mu}dx^{\nu},
\end{equation}
where $\Phi({\bf{r}})\ll1$ is the gravitational potential of the FDM soliton and $\Psi({\bf{r}})\ll1$ the corresponding perturbation of the spatial curvature, \Mov{illustrated }
in Fig.~\ref{fig:soliton}. Consider the linear perturbation $h_{\mu\nu}$ in the background metric tensor $g^{(\rm B)}_{\mu\nu}$ as
\begin{equation}
g_{\mu\nu} = g^{(\rm B)}_{\mu\nu} + h_{\mu\nu}.
\end{equation}
Under the transverse traceless Lorentz gauge condition 
$\nabla_{\mu}h^{\mu\nu}=0$ and $g^{(\rm B)\,\mu\nu}h_{\mu\nu}=0$\Mov{, }%
the propagation equation for GWs $h_{\mu\nu}$\Mov{ follows}
\begin{equation}
\nabla^2h_{ij}
+\nabla(\Phi-\Psi)\cdot\nabla h_{ij}-\frac{1-2\Phi-2\Psi}{c^2}\frac{\partial^2 h_{ij}}{\partial t^2}=0,  
\end{equation}
where we have neglected the higher-order nonlinear terms \Mov{following}
~\cite{Peters:1974gj}
. Using the eikonal approximation~\cite{Baraldo:1999ny}, the GW \Mov{perturbation }tensor can be represented as 
\begin{equation}
h_{ij} = u e_{ij},
\end{equation}
where $e_{ij}$ is the polarization tensor of the GW and $u$ is a scalar wave. Since the change in the polarization tensor by gravitational lensing is on the order of $\Phi({\bf{r}})\ll1$, we assume that the polarization tensor does not change during the propagation of GW. 
Thus, we obtain the propagation equation for the scalar wave as
\begin{equation}
\label{eq:gw}
\nabla^2u+\nabla(\Phi-\Psi)\cdot\nabla u-\frac{1-2\Phi-2\Psi}{c^2}\frac{\partial^2 u}{\partial t^2}=0,  
\end{equation}
We can rewrite it as follows
\begin{equation}
a^2\nabla^2u+a^2\nabla b\cdot\nabla u-\frac{\partial^2u}{\partial t^2}=0,\label{eq:SolitonWave}
\end{equation}
where the effective \Mov{wave }speed 
$a$ and parameter $b$ are defined as
\begin{eqnarray}
\frac{1-2\Phi-2\Psi}{c^2}&=&\frac{1}{a^2},\\
\Phi-\Psi&=&b. \label{eq:bDef}
\end{eqnarray}

There are several methods to solve the above equation in a bounded domain $\Omega\subset\mathbb{R}^3$ with boundary $\partial\Omega$ from $t=0$ to $t=T$. Here, we will use the finite element method to solve the weak form of the above equation\Mov{\footnote{\Mov{Given the specific wave equation \eqref{eq:SolitonWave}, the calculations of Eqs.~(\ref{eq:Phi.v}-\ref{eq:MijDef}) just follow \cite{He:2021hhl,Qiu:2022dya,He:2022sjf,Yin:2023kzr} step by step, with some modifications. 
}}}
\begin{eqnarray}
\left(\phi,\frac{\partial u}{\partial t}\right)_{\Omega} &=&(\phi,v)_{\Omega},\label{eq:Phi.v}\\
\nonumber
\left(\phi,\frac{\partial v}{\partial t}\right)_{\Omega}&=&-\left(\nabla(a^2\phi),\nabla u\right)_{\Omega}-\left(a\phi,\frac{\partial u}{\partial t}\right)_{\partial\Omega}\\
&&+\left((a^2\phi)\nabla b,\nabla u\right)_{\Omega},
\end{eqnarray}
where $\phi$ is a test function\Mov{,  $v\equiv\frac{\partial u}{\partial t}$} and $(f,g)_{\Omega}=\int_{\Omega}f({\bf{x}})g({\bf{x}})d{\bf{x}}$ is \Mov{the $\mathcal{L}^2$ inner product of $f$ and $g$
}
. In the second equality, we have imposed an absorbing boundary condition $\hat{n}\cdot\nabla u=-\frac{1}{a}\frac{\partial u}{\partial t}$ on $\partial\Omega\times(0,T]$. It should be noted that the term $-\left(a\phi,\frac{\partial u}{\partial t}\right)_{\partial\Omega}$ will be neglected for the special boundary face from which GWs enter our simulation domain. First, we turn to the time discretization as Rothe's method
\begin{eqnarray}
\left(\phi,\frac{u^n-u^{n-1}}{k}\right)_{\Omega}&=&\left(\phi,\theta v^n+(1-\theta)v^{n-1}\right)_{\Omega},\\
\nonumber
\left(\phi,\frac{v^n-v^{n-1}}{k}\right)_{\Omega}&=&-\left(\nabla(a^2\phi),\nabla[\theta u^n+(1-\theta)u^{n-1}]\right)_{\Omega}\\
\nonumber
&&+\left((a^2\phi)\nabla b,\nabla[\theta u^n+(1-\theta)u^{n-1}]\right)_{\Omega}\\
&&-\left(a\phi,\frac{u^n-u^{n-1}}{k}\right)_{\partial\Omega},
\end{eqnarray}  
where a superscript $n$ indicates the number of 
time step\Mov{s}, $k=t_n-t_{n-1}$ is the length of the \Mov{current }
time step and $\theta = \frac{1}{2}$ is the choice of the Crank-Nicolson method. For clarity, we relate the 
solution $u^n$ and its time derivative $v^n$ at time $t_n$ to the solution $u^{n-1}$ and $v^{n-1}$ at the previous time step $t_{n-1}$ as
\begin{multline}
\left(\phi,u^{n}\right)_{\Omega}+k^{2}\theta^{2}\left([\nabla(a^{2}\phi)-(a^{2}\phi)\nabla b],\nabla u^{n}\right)_{\Omega}\\
+k\theta\left(a\phi,u^{n}\right)_{\partial\Omega}=\left(\phi,u^{n-1}\right)_{\Omega}\\
-k^{2}\theta(1-\theta)\left([\nabla(a^{2}\phi)-(a^{2}\phi)\nabla b],\nabla u^{n-1}\right)_{\Omega},
\end{multline}\vspace{-.8cm}
\begin{multline}
\left(\phi,v^{n}\right)_{\Omega}=\left(\phi,v^{n-1}\right)_{\Omega}-k\theta\left([\nabla(a^{2}\phi)-(a^{2}\phi)\nabla b],\nabla u^{n}\right)_{\Omega}\\
-\left(a\phi,u^{n}\right)_{\partial\Omega}-k(1-\theta)\left([\nabla(a^{2}\phi)-(a^{2}\phi)\nabla b],\nabla u^{n-1}\right)_{\Omega}\\
+\left(a\phi,u^{n-1}\right)_{\partial\Omega}.
\end{multline}
The next step is space discretization using the usual finite element method
. At each time step, we use the same 
shape 
functions $\phi_i$ 
to approximate $u^n$, $v^n$, $u^{n-1}$ and $v^{n-1}$ as
\begin{eqnarray}
u^n&\approx&\sum_iU^n_i\phi_i,\label{eq:uDecomp}\\
v^n&\approx&\sum_iV^n_i\phi_i,\\
u^{n-1}&\approx&\sum_iU^{n-1}_i\phi_i,\\
v^{n-1}&\approx&\sum_iV^{n-1}_i\phi_i.
\end{eqnarray}
\Mov{We finally obtain }
the following linear system
\begin{multline}
\left[\mathcal{M}+k^{2}\theta^{2}(\mathcal{A}+\mathcal{D}-\mathcal{C})+k\theta\mathcal{B}\right]U^{n}=\\
\left[\mathcal{M}-k^{2}\theta(1-\theta)(\mathcal{A}+\mathcal{D}-\mathcal{C})+k\theta\mathcal{B}\right]U^{n-1}\\
+k\mathcal{M}V^{n-1}\,,\label{eq:u}
\end{multline}\vspace{-.9cm}
\begin{multline}
\left[\mathcal{M}+k^{2}\theta^{2}(\mathcal{A}+\mathcal{D}-\mathcal{C})+k\theta\mathcal{B}\right]V^{n}=\\
\left[\mathcal{M}-k^{2}\theta(1-\theta)(\mathcal{A}+\mathcal{D}-\mathcal{C})-k(1-\theta)\mathcal{B}\right]V^{n-1}\\
-k(\mathcal{A}+\mathcal{D}-\mathcal{C})U^{n-1}\,,\label{eq:v}
\end{multline}
where the \Mov{matrix }elements 
are defined as 
\begin{eqnarray}
\mathcal{A}_{ij}&=&\left(a^2\nabla\phi_i,\nabla\phi_j\right)_{\Omega},\\
\mathcal{B}_{ij}&=&\left(a\phi_i,\phi_j\right)_{\partial\Omega},\\
\mathcal{C}_{ij}&=&\left(a^2\nabla(b)\phi_i,\nabla\phi_j\right)_{\Omega},\\
\mathcal{D}_{ij}&=&\left(\nabla(a^2)\phi_i,\nabla\phi_j\right)_{\Omega},\\
\mathcal{M}_{ij}&=&\left(\phi_i,\phi_j\right)_{\Omega}.\label{eq:MijDef}
\end{eqnarray}

\section{Numerical Simulation}
\label{sec:sim}
\Mov{We then proceed to }
solve the coupled system 
Eqs.~(\ref{eq:u}\Mov{,}
\ref{eq:v}) by a modified $\mathtt{GWsim}$~\cite{He:2021hhl} code, which is further based on the publicly available finite element package $\mathtt{deal.ii}$~\cite{dealii}. 
In detail, we simulate the propagation of sinusoidal plane GWs with amplitude $A=1$ and frequency $f\sim10^{-8}{\rm~Hz}$ through a cylinder with a radius of $7.5{\rm~pc}$ and a length of $15{\rm~pc}$. The cylinder axis is taken along the $x$-axis ranging from $-7.5{\rm~pc}$ to $7.5{\rm~pc}$. The incident GWs travel along the $x$-axis. Although GWs can sweep the cylinder in $15{\rm~pc/c}$, the simulations 
last\Mov{ed} 
$22.575{\rm~pc/c}$.
\Mov{For a }
simulation domain \Mov{with }
refinement of $2^8$\Mov{,} with a total of $1.7\times 10^8$ degrees of freedom (or nodal points), each simulation \Mov{requires }
$320$ CPU cores and about $14{\rm~k}$ CPU hours.
To investigate the effect of the FDM soliton on GW propagation, at the center, we locate an FDM soliton with a radius of $5.8{\rm~pc}$, $4.4{\rm~pc}$ or $3.5{\rm~pc}$\Mov{, respectively,} \Mov{made of }
FDM particles with mass $m=8\times10^{-21}{\rm~eV/c^2}$, $m=10\times10^{-21}{\rm~eV/c^2}$ or $m=12\times10^{-21}{\rm~eV/c^2}$\Mov{,} respectively. Their gravitational potential and the corresponding perturbation of the spatial curvature are shown in Fig.~\ref{fig:soliton}. 

\begin{figure}[]
\begin{center}
\includegraphics[width= 1\columnwidth]{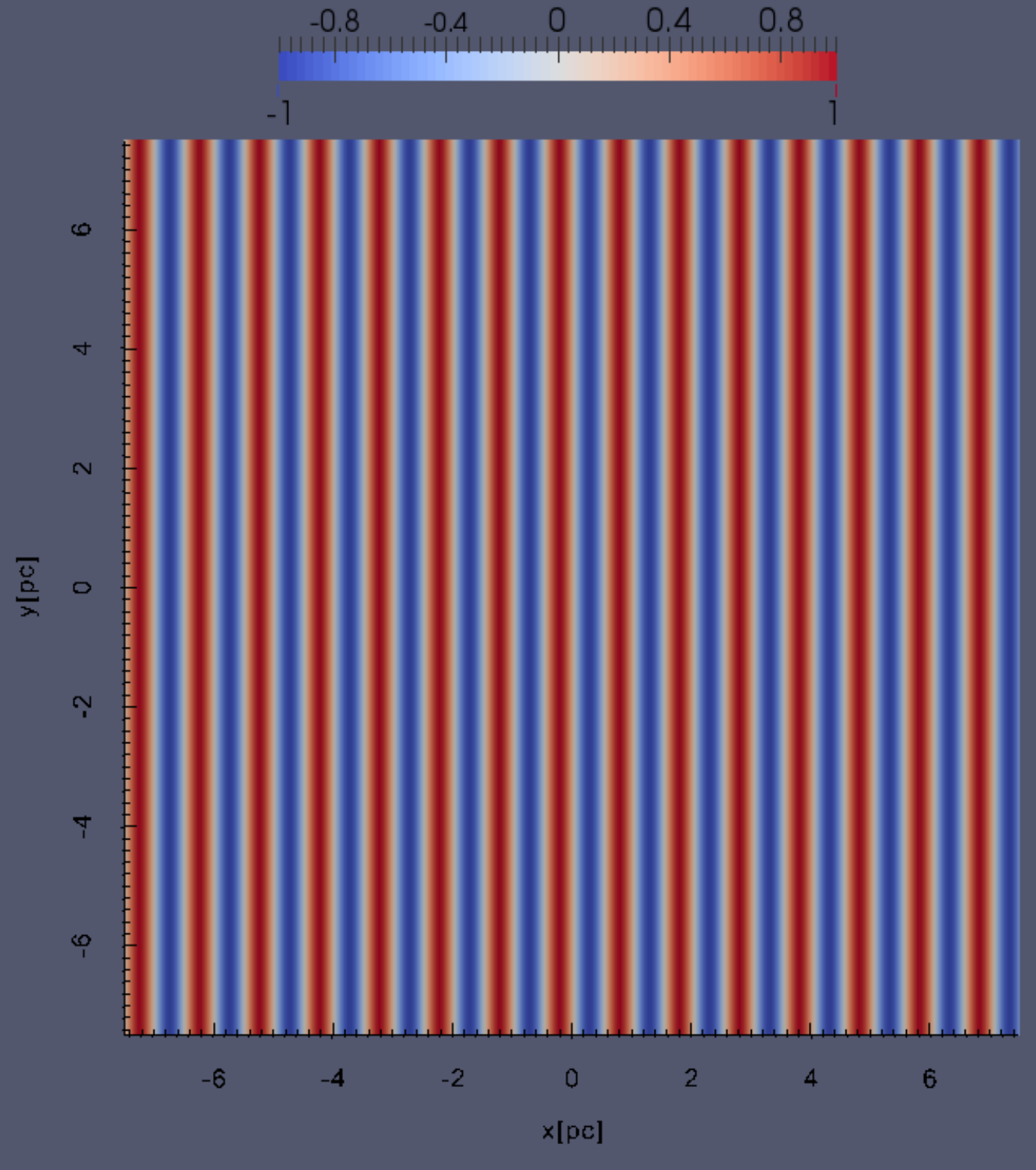}
\end{center}
\caption{Last snapshots of \Ke{$u$ in the }$x$-$y$ plane at $z=0$ for simulations with FDM soliton with $m=12\times10^{-21}{\rm~eV/c^2}$. \Ke{This figure, compared with Fig.~\ref{fig:xyl}, illustrates that due to the gravitational potential flatness of the FDM soliton, any gravitational lensing effects cannot be distinguished in the $x$-$y$ plane, see text.}
}\label{fig:xy}
\end{figure}

Since GWs with $f\sim10^{-8}{\rm~Hz}$ have a wavelength of $1{\rm~pc}$, \Mov{such GWs complete }
$15$ 
periods in the simulation domain for the \Mov{case }
without FDM soliton, as shown in \Ke{Fig.~\ref{fig:xyl}, in Appendix~\ref{sec:GWprop}}. However, due to the extremely flat gravitational potential $\sim 10^{-6}$ of the FDM soliton with $m=12\times10^{-21}{\rm~eV/c^2}$, 
any gravitational lensing effect, such as the Shapiro time delay,\Mov{ remains very faint and cannot be observed in the propagation direction ($x$-$y$ plane) as seen} in  Fig.~\ref{fig:xy}. 

\begin{figure*}[]
\begin{center}
\includegraphics[width= 17cm]{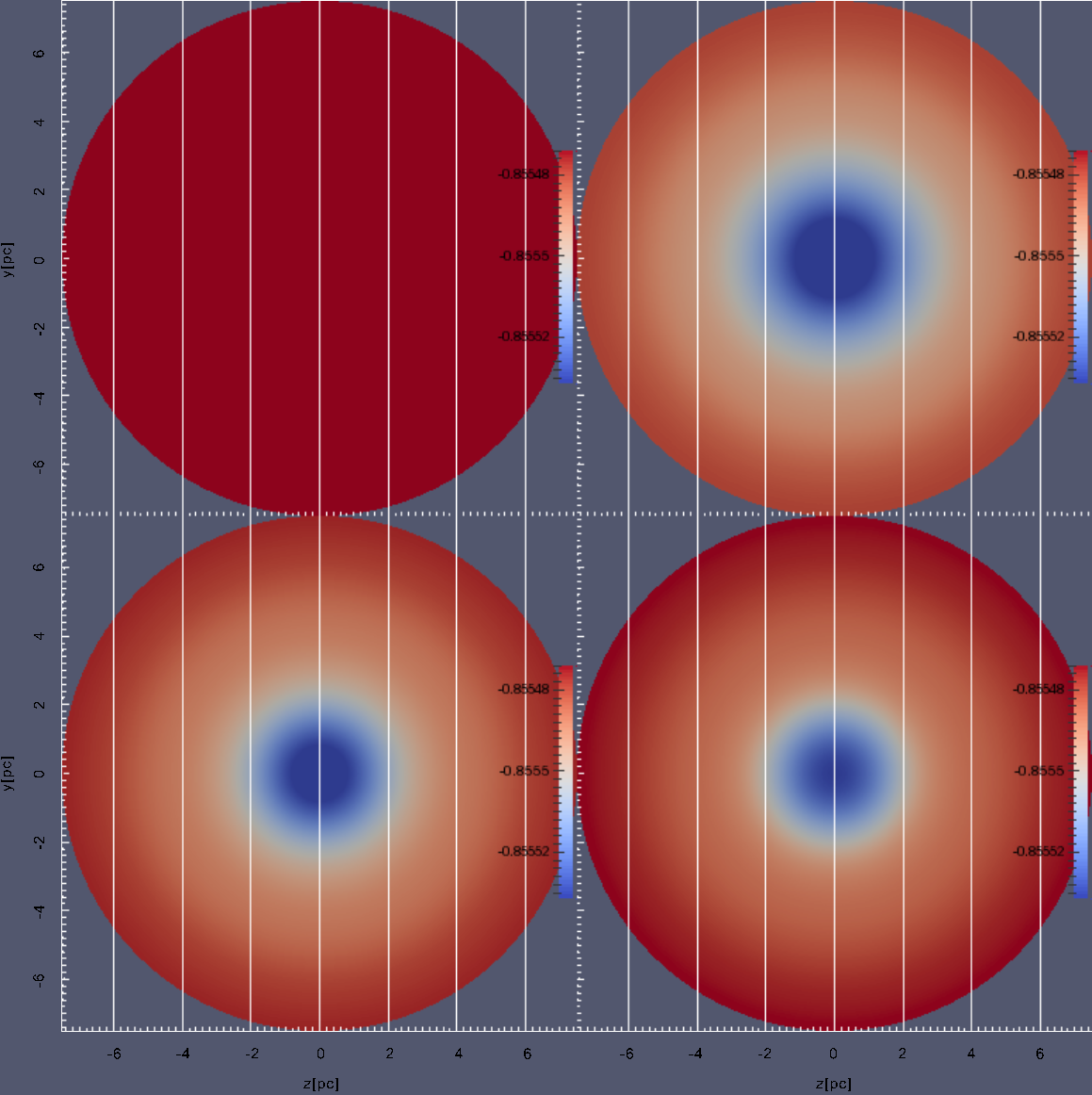}
\end{center}
\caption{Last snapshots \Ke{of $u$ in the }$y$-$z$ plane with $r=7.5{\rm~pc}$ at $x=7.25{\rm~pc}$ for simulations without FDM soliton (upper left) and with FDM soliton with $m=8\times10^{-21}{\rm~eV/c^2}$ (upper right), $m=10\times10^{-21}{\rm~eV/c^2}$ (lower left) and $m=12\times10^{-21}{\rm~eV/c^2}$ (lower right) respectively.}
\label{fig:yz}
\end{figure*}

Fortunately, the gravitational lensing magnification is observable in the snapshot of the $y$-$z$ plane, as shown in Fig.~\ref{fig:yz}. \Ke{It displays the scalar wave magnitude $u$ in a different colour scale than Fig.~\ref{fig:xy}, that allows to visualise the magnification of GWs induced by the soliton. Magnification can be defined as $F(x)=\left|\frac{u(x)-u_{\rm v}(x)
}{u_{\rm v}(x)}\right|$,\footnote{\Ke{Note that our definition differs from Eq.~(9) of~\cite{Takahashi:2003ix} as our signal is relatively small. This definition therefore allows to tackle tiny lensing effects.}} where $u_{\rm v}$ is the unlensed, vacuum, value of the scalar wave magnitude. }Compared with the simulation without FDM soliton (upper left snapshot), the simulations with FDM soliton (the other three snapshots) feature \Mov{clear }gravitational lensing magnification.
The amplitude of GWs is magnified by the FDM soliton by at most $10^{-4}$. Although this magnification factor is very small, its corresponding magnification zone \Mov{extends to astrophysical scales. For example, in Fig.~\ref{fig:yz}, the blue part of the magnification zones extends to }
a size of $6{\rm~pc}$ for $m=8\times10^{-21}{\rm~eV/c^2}$ (upper right), $5{\rm~pc}$ for $m=10\times10^{-21}{\rm~eV/c^2}$ (lower left) and $4{\rm~pc}$ for $m=12\times10^{-21}{\rm~eV/c^2}$ (lower right), respectively.
The lensed GWs \Mov{then propagate freely }
as almost plane waves \Mov{after passing }
the FDM soliton\Mov{ region}. \Mov{Passed that region, }
the magnification factor and the \Mov{extent of the  }magnification 
zone \Mov{remain }
unchanged
.

\section{Summary and discussion}
\label{sec:sum} 
In this paper, we introduce a potential detection method for FDM with mass $\sim10^{-20}\rm~{eV}/c^2$. First, we briefly review the calculation of the FDM soliton 
density profiles\Mov{, }
their corresponding gravitational potential and perturbation of the spatial curvature for \Mov{a set of FDM particle masses ranging }$m=8\times10^{-21}{\rm~eV/c^2}$, $m=10\times10^{-21}{\rm~eV/c^2}$ and $m=12\times10^{-21}{\rm~eV/c^2}$, respectively. The difference between the gravitational potential and the perturbation of the spatial curvature inside the FDM soliton characterizes the propagation equation for GWs. \Mov{We then }
solve this propagation equation\Mov{ using a finite element method, transforming }
it into a matrix form
. \Mov{This allows us to }
simulate the gravitational lensing effects of FDM solitons on GWs with $f\sim10^{-8}{\rm~Hz}$ \Mov{for an }
FDM particle mass \Mov{range of }
$m=8\times10^{-21}{\rm~eV/c^2}$, $m=10\times10^{-21}{\rm~eV/c^2}$ and $m=12\times10^{-21}{\rm~eV/c^2}$, respectively. Although we cannot observe the Shapiro time delay due to the extremely flat gravitational potential $\sim 10^{-6}$, we can observe the gravitational lensing magnification. More precisely, a 
magnification factor \Mov{of at most $10^{-4}$ }occurs in a 
magnification zone\Mov{ of astrophysical size}: $6{\rm~pc}$ for $m=8\times10^{-21}{\rm~eV/c^2}$, $5{\rm~pc}$ for $m=10\times10^{-21}{\rm~eV/c^2}$ and $4{\rm~pc}$ for $m=12\times10^{-21}{\rm~eV/c^2}$, respectively.

GWs with $f\sim10^{-8}{\rm~Hz}$ from all directions would \Mov{contribute to the low-frequency }
GW background, which 
PTA\Mov{ is expected to detect}~\cite{NANOGrav:2023gor,EPTA:2023fyk}.
\Mov{In the case of }
FDM with $m\sim10^{-23}{\rm~eV/c^2}$
\Mov{, a similar oscillation signature in }
PTA measurements as the GW background\Mov{ is expected}~\cite{Khmelnitsky:2013lxt}, \Mov{while the case of }FDM with $m\sim10^{-20}{\rm~eV/c^2}$\Mov{, the corresponding oscillation signal} should be out of the PTA detection band due to its higher oscillation frequency.
\Mov{Nevertheless}
, the FDM soliton lens in the center of our galaxy \Mov{still }introduces another possibility of detection of FDM with $m\sim10^{-20}{\rm~eV/c^2}$ by PTA.
More precisely, since the magnification zone size is $\sim 10{\rm~pc}$, 
comparable to the \Mov{PTA's }arm length\Mov{, 
$\sim1{\rm~kpc}$}, \Mov{it should induce }
an antisotropy of $\sim10^{-4}$ over an large enough solid angle in the GW background in the direction of the FDM soliton\Mov{, see Fig.~\ref{fig:PTAsourceAngle}\begin{figure}[ht]
\centering
\includegraphics[width=\linewidth]{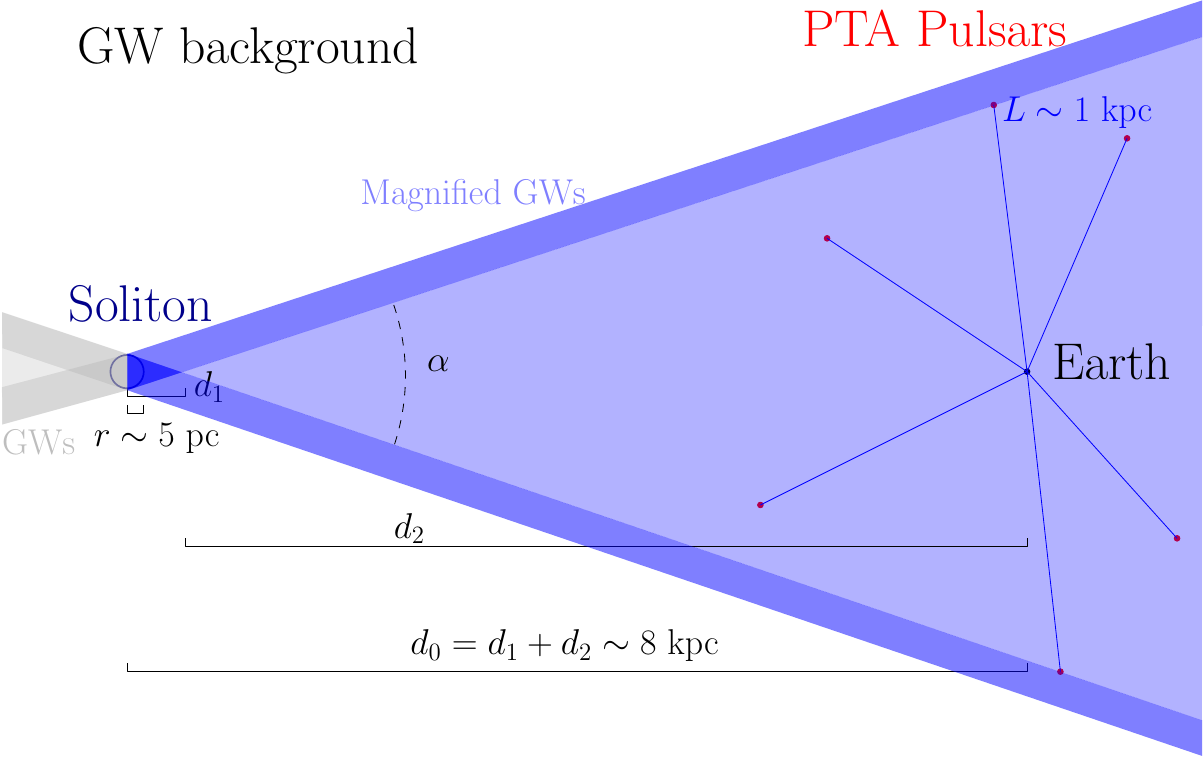}
\caption{\Mov{Formation of the nHz GW background anisotropy induced by the lensing effect of the Galactic Center soliton. The anisotropy proceeds from the maximum angle of magnified GWs sweeping the PTA.}}
\label{fig:PTAsourceAngle}
\end{figure}. For example, the 68 pulsars of the NANOGrav 15-year dataset, located at distances of order $1{\rm~kpc}$ \cite{NANOGrav:2023hde} from Earth, form a multiple arms GW detector located at a distance of order $8{\rm~kpc}$ \cite{Nesti:2013uwa} from the soliton that magnifies the GW background. The signal size given by the soliton magnification zone covers $10{\rm~pc}$ (Fig.~\ref{fig:yz}). From the illustration of Fig.~\ref{fig:PTAsourceAngle}, we can compute that \begin{align}
     \ell = \frac{\pi}{\alpha} \simeq & \frac{\pi}{2\arctan\left(\frac{r+L}{d_0}\right)}\sim 12.
\end{align}
This could be extrapolated to predict an extra point in \Ke{the angular power spectrum represented in }Fig.~4 of \cite{NANOGrav:2023tcn} at $C_\ell/C_0 \sim 10^{-4}$ for $\ell\sim 12$.\Ke{ Recall that the full sky angular power spectrum gives the fluctuation power $\ell (2\ell +1) C_\ell$ at a given angular separation $\pi/\ell$.} }Of course, \Mov{this is still out of detection range, $<20\%$, of present day PTAs}
~\cite{NANOGrav:2023tcn}.
\Ke{

Although the presence in the middle of the Milky Way of the Sgr~A$^*$ SuperMassive BH of mass $M_{\rm SMBH} \sim 10^6 M_\odot$ should modify the formation of the soliton~\cite{Davies:2019wgi,Tan:2024dne}, its effect on the density profile of the FDM soliton can be neglected when the halo mass is $M_{\rm{halo}}=1\times 10^{12} ~M_{\odot}$~\cite{Wang:2019ubx} while the FDM mass is $\sim10^{-20}\rm~{eV}/c^2$ , as seen in Fig.~2 of \cite{Davies:2019wgi}.

As for its 
gravitational lensing effect, Sgr~A$^*$ 
acts as a strong gravitational lens. However, this 
effect 
only dominates when Sgr~A$^*$ receives GWs with frequency much higher than $f\sim10^{-8}{\rm~Hz}$ or with wavelength much shorter than $1{\rm~pc}$. Therefore, the lensing effects of Sgr~A$^*$ can be neglected here.

In 
this paper, we only study the lensing effects of FDM soliton of the Milky Way. In this case, the detectors are located in the NFW-like halo outside of this FDM soliton. The observed GWs from every direction are lensed by the surrounding FDM density, including that of the FDM soliton and the NFW-like halo density. Obviously, the lensing effects observed from the FDM soliton direction dominates over the lensing effects observed in any other direction because of the large density ratio between the direction of the FDM soliton, located in the galactic centre, and any other direction. 
Furthermore, the FDM soliton potential is so flat (Fig.~\ref{fig:soliton}) that the Shapiro time delay can't be observed (Fig.~\ref{fig:xy}) and its amplification factor is tiny $\sim10^{-4}$ (Fig.~\ref{fig:yz}). This is all the more the case for the much flatter potential of the much lower NFW-like halo density outside of this FDM soliton.

}
\vspace{5mm}
\noindent {\bf Acknowledgments}
We acknowledge the use of HPC Cluster of Tianhe II in National Supercomputing Center in Guangzhou. Ke Wang is supported by grants from the National Key Research and Development Program of China (grant No. 2021YFC2203003).
MLeD acknowledges the financial support by the Lanzhou University starting fund, the Fundamental Research Funds for the Central Universities (Grants No. lzujbky-2019-25 and lzujbky-2025-jdzx07), the Natural Science Foundation of Gansu Province (No. 22JR5RA389 and No.25JRRA799), National Science Foundation of China  (NSFC grant No.12247101)
and the ‘111 Center’ under Grant No. B20063.

\appendix
\section{GW propagation in vacuum}\label{sec:GWprop}
Consider the linear perturbation $h_{\mu\nu}$ in the vacuum background metric tensor $g^{(\rm B)}_{\mu\nu}$, following~\cite{He:2021hhl,Qiu:2022dya,He:2022sjf,Yin:2023kzr}, as
\begin{equation}
g_{\mu\nu} = g^{(\rm B)}_{\mu\nu} + h_{\mu\nu}.
\end{equation}
Under the transverse traceless Lorentz gauge condition 
$\nabla_{\mu}h^{\mu\nu}=0$ and $g^{(\rm B)\,\mu\nu}h_{\mu\nu}=0$, %
the \Ke{vacuum }propagation equation for GWs $h_{\mu\nu}$ follows
\begin{equation}
\nabla^2h_{ij}
-\frac{1}{c^2}\frac{\partial^2 h_{ij}}{\partial t^2}=0,  
\end{equation}
where 
the higher-order nonlinear terms \Ke{have been neglected }following
~\cite{Peters:1974gj}
. \Ke{The }
eikonal approximation~\cite{Baraldo:1999ny}\Ke{ allows one to represent }
the GW perturbation tensor 
as 
\begin{equation}
h_{ij} = u e_{ij},
\end{equation}
\Ke{with }
the polarization tensor \Ke{noted  $e_{ij}$ }
and \Ke{the }
scalar wave\Ke{, $u$}. 
\Ke{The scalar wave }
propagation equation \Ke{then reads}
\begin{equation}
\nabla^2u-\frac{1}{c^2}\frac{\partial^2 u}{\partial t^2}=0.\label{eq:AppWave}  
\end{equation}

\Ke{We will use the finite element method }
to solve the above equation in a bounded domain $\Omega\subset\mathbb{R}^3$ with boundary $\partial\Omega$ from $t=0$ to $t=T$. Here, we will \Ke{focus on }
the weak form of the above equation
\begin{align}
    &\int_\Omega c^2\phi\nabla^2 u{\rm~d}x-\frac{\partial^2}{\partial t^2}\int_{\Omega}\phi u{\rm~d}x=0,\\
\nonumber
&\int_\Omega c^2\phi\nabla^2 u{\rm~d}x\\
&=-\int_\Omega\nabla(c^2\phi)\cdot\nabla u{\rm~d}x+\int_{\partial\Omega}(c^2\phi)(\hat{n}\cdot\nabla u){\rm~d}x,
\label{eq:AppGreen}
\end{align}
where we have multiplied Eq.~(\ref{eq:AppWave}) by is a test function $\phi$, integrate\Ke{d} over $\Omega$ and used the Green’s formula in Eq.~(\ref{eq:AppGreen}).
With definitions $v\equiv\frac{\partial u}{\partial t}$ and $(f,g)_{\Omega}\equiv\int_{\Omega}f({\bf{x}})g({\bf{x}})d{\bf{x}}$, the $\mathcal{L}^2$ inner product of $f$ and $g$, we have
\begin{eqnarray}
\left(\phi,\frac{\partial u}{\partial t}\right)_{\Omega} &=&(\phi,v)_{\Omega},\\
\left(\phi,\frac{\partial v}{\partial t}\right)_{\Omega}&=&-\left(\nabla(c^2\phi),\nabla u\right)_{\Omega}-\left(c\phi,\frac{\partial u}{\partial t}\right)_{\partial\Omega},\label{eq:2ndEq}
\end{eqnarray}
where, in \Ke{Eq.~\ref{eq:2ndEq}}
, \Ke{the }
absorbing boundary condition $\hat{n}\cdot\nabla u=-\frac{1}{c}\frac{\partial u}{\partial t}$ \Ke{is set at }
$\partial\Omega\times(0,T]$. \Ke{Note }
that the term $-\left(c\phi,\frac{\partial u}{\partial t}\right)_{\partial\Omega}$\Ke{, in the boundary section }
from which GWs enter our simulation domain\Ke{, is neglected}. 

\Ke{The }
finite element
method's 
time discretization \Ke{uses }
Rothe's method
\begin{eqnarray}
\left(\phi,\frac{u^n-u^{n-1}}{k}\right)_{\Omega}&=&\left(\phi,\theta v^n+(1-\theta)v^{n-1}\right)_{\Omega},\\
\nonumber
\left(\phi,\frac{v^n-v^{n-1}}{k}\right)_{\Omega}&=&-\left(\nabla(c^2\phi),\nabla[\theta u^n+(1-\theta)u^{n-1}]\right)_{\Omega}\\
&&-\left(c\phi,\frac{u^n-u^{n-1}}{k}\right)_{\partial\Omega},
\end{eqnarray}  
where \Ke{the }
superscript $n$ indicates the number of 
time steps, \Ke{the current time step length is }$k=t_n-t_{n-1}$ 
and $\theta = \frac{1}{2}$ \Ke{corresponds to }
the choice of the Crank-Nicolson method. \Ke{Expanding}
, we relate the 
solution $u^n$ and its time derivative $v^n$ at time\Ke{ step} $t_n$ to the\Ke{ir values }
at the previous time step $t_{n-1}$ \Ke{following}
\begin{multline}
\left(\phi,u^{n}\right)_{\Omega}+k^{2}\theta^{2}\left(\nabla(c^{2}\phi),\nabla u^{n}\right)_{\Omega}+k\theta\left(c\phi,u^{n}\right)_{\partial\Omega}\\=\left(\phi,u^{n-1}\right)_{\Omega}
-k^{2}\theta(1-\theta)\left(\nabla(c^{2}\phi),\nabla u^{n-1}\right)_{\Omega},
\end{multline}\vspace{-.8cm}
\begin{multline}
\left(\phi,v^{n}\right)_{\Omega}\\=\left(\phi,v^{n-1}\right)_{\Omega}-k\theta\left(\nabla(c^{2}\phi),\nabla u^{n}\right)_{\Omega}
-\left(c\phi,u^{n}\right)_{\partial\Omega}\\-k(1-\theta)\left(\nabla(c^{2}\phi),\nabla u^{n-1}\right)_{\Omega}
+\left(c\phi,u^{n-1}\right)_{\partial\Omega}.
\end{multline}

\Ke{Space }
discretization\Ke{, following }
the usual finite element method
\Ke{, uses }
the same 
shape 
functions $\phi_i$ \Ke{at each time step }
to approximate $u^n$, $v^n$, $u^{n-1}$ and $v^{n-1}$ as
\begin{eqnarray}
u^n&\approx&\sum_iU^n_i\phi_i,\\
v^n&\approx&\sum_iV^n_i\phi_i,\\
u^{n-1}&\approx&\sum_iU^{n-1}_i\phi_i,\\
v^{n-1}&\approx&\sum_iV^{n-1}_i\phi_i.
\end{eqnarray}
\Ke{The solution proceeds then from }
the following linear system
\begin{multline}
\left[\mathcal{M}+k^{2}\theta^{2}(\mathcal{A}+\mathcal{D})+k\theta\mathcal{B}\right]U^{n}=\\
\left[\mathcal{M}-k^{2}\theta(1-\theta)(\mathcal{A}+\mathcal{D})+k\theta\mathcal{B}\right]U^{n-1}
+k\mathcal{M}V^{n-1},
\label{eq:Appu}
\end{multline}\vspace{-.9cm}
\begin{multline}
\left[\mathcal{M}+k^{2}\theta^{2}(\mathcal{A}+\mathcal{D})+k\theta\mathcal{B}\right]V^{n}=\\
\left[\mathcal{M}-k^{2}\theta(1-\theta)(\mathcal{A}+\mathcal{D})
-k(1-\theta)\mathcal{B}\right]V^{n-1}\\
-k(\mathcal{A}+\mathcal{D})U^{n-1},
\label{eq:Appv}
\end{multline}
where the matrix elements 
are defined as 
\begin{eqnarray}
\mathcal{A}_{ij}&=&\left(c^2\nabla\phi_i,\nabla\phi_j\right)_{\Omega},\\
\mathcal{B}_{ij}&=&\left(c\phi_i,\phi_j\right)_{\partial\Omega},\\
\mathcal{D}_{ij}&=&\left(\nabla(c^2)\phi_i,\nabla\phi_j\right)_{\Omega},\\
\mathcal{M}_{ij}&=&\left(\phi_i,\phi_j\right)_{\Omega}.
\end{eqnarray}

\begin{figure}[]
\begin{center}
\includegraphics[width= 1\columnwidth]{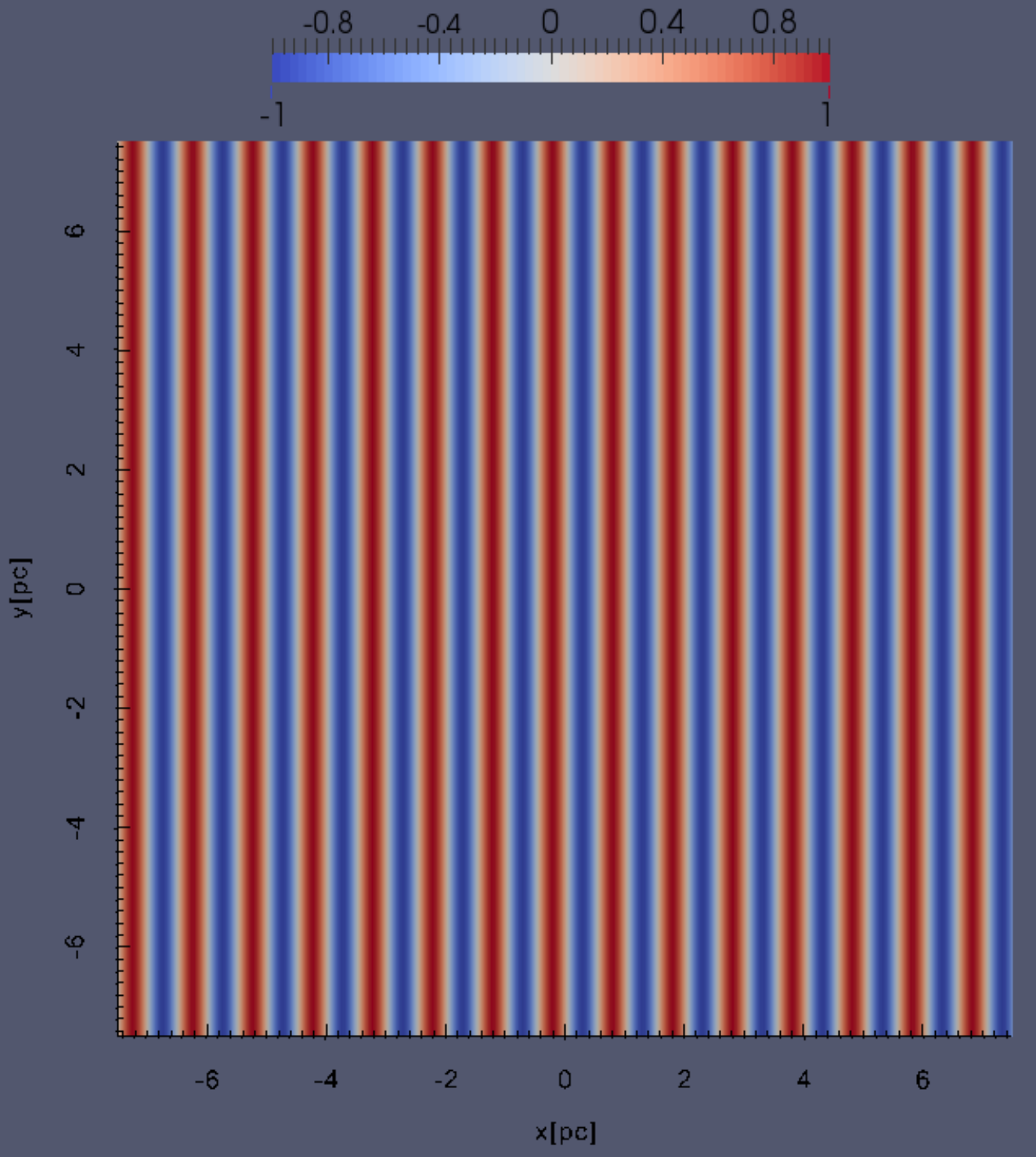}
\end{center}
\caption{\Ke{Last snapshots of $u$ in the $x$-$y$ plane at $z=0$ for simulations in vacuum. 
}}
\label{fig:xyl}
\end{figure}

The coupled system Eqs.~(\ref{eq:Appu},\ref{eq:Appv}) can be numerically solved by \Ke{the }$\mathtt{GWsim}$~\cite{He:2021hhl} code, \Ke{itself }
based on the publicly available finite element package $\mathtt{deal.ii}$~\cite{dealii}. 
In \Ke{this simple example}
, we simulate the propagation of sinusoidal plane GWs with amplitude $A=1$ and frequency $f\sim10^{-8}{\rm~Hz}$ through a cylinder \Ke{of }
radius 
$7.5{\rm~pc}$ and 
length 
$15{\rm~pc}$. The cylinder axis is \Ke{set }
along the $x$-axis\Ke{, with values} ranging from $-7.5{\rm~pc}$ to $7.5{\rm~pc}$. The incident GWs travel along the $x$-axis. \Ke{While }
GWs 
sweep the cylinder in $15{\rm~pc/c}$, the simulations lasted $22.575{\rm~pc/c}$.
For a simulation domain with refinement 
$2^8$, 
a total of $1.7\times 10^8$ degrees of freedom (or nodal points), each simulation \Ke{consumes }
$320$ CPU cores and about $14{\rm~k}$ CPU hours.
\Ke{As the wavelength of }
GWs with $f\sim10^{-8}{\rm~Hz}$ \Ke{corresponds to }
$1{\rm~pc}$, \Ke{they }
complete $15$ periods in the simulation domain, as shown in Fig.~\ref{fig:xyl}.




\begin{thebibliography}{99}
\frenchspacing

\bibitem{Rubin:1982kyu}
V.~C.~Rubin, W.~K.~Ford, Jr., N.~Thonnard and D.~Burstein,
``Rotational properties of 23 SB galaxies,''
Astrophys. J. \textbf{261}, 439 (1982)

\bibitem{Davis:1985rj}
M.~Davis, G.~Efstathiou, C.~S.~Frenk and S.~D.~M.~White,
``The Evolution of Large Scale Structure in a Universe Dominated by Cold Dark Matter,''
Astrophys. J. \textbf{292}, 371-394 (1985)

\bibitem{Clowe:2006eq}
D.~Clowe, M.~Bradac, A.~H.~Gonzalez, M.~Markevitch, S.~W.~Randall, C.~Jones and D.~Zaritsky,
``A direct empirical proof of the existence of dark matter,''
Astrophys. J. Lett. \textbf{648}, L109-L113 (2006)
[arXiv:astro-ph/0608407 [astro-ph]].

\bibitem{Sanders:2006sz}
R.~H.~Sanders,
``Modified gravity without dark matter,''
Lect. Notes Phys. \textbf{720}, 375-402 (2007)
[arXiv:astro-ph/0601431 [astro-ph]].

\bibitem{Planck:2018vyg}
N.~Aghanim \textit{et al.} [Planck],
``Planck 2018 results. VI. Cosmological parameters,''
Astron. Astrophys. \textbf{641}, A6 (2020)
[erratum: Astron. Astrophys. \textbf{652}, C4 (2021)]
[arXiv:1807.06209 [astro-ph.CO]].

\bibitem{PandaX-II:2016vec}
A.~Tan \textit{et al.} [PandaX-II],
``Dark Matter Results from First 98.7 Days of Data from the PandaX-II Experiment,''
Phys. Rev. Lett. \textbf{117}, no.12, 121303 (2016)
[arXiv:1607.07400 [hep-ex]].

\bibitem{LUX:2015abn}
D.~S.~Akerib \textit{et al.} [LUX],
``Improved Limits on Scattering of Weakly Interacting Massive Particles from Reanalysis of 2013 LUX Data,''
Phys. Rev. Lett. \textbf{116}, no.16, 161301 (2016)
[arXiv:1512.03506 [astro-ph.CO]].

\bibitem{LUX:2016ggv}
D.~S.~Akerib \textit{et al.} [LUX],
``Results from a search for dark matter in the complete LUX exposure,''
Phys. Rev. Lett. \textbf{118}, no.2, 021303 (2017)
[arXiv:1608.07648 [astro-ph.CO]].

\bibitem{Carr:2016drx}
B.~Carr, F.~Kuhnel and M.~Sandstad,
``Primordial Black Holes as Dark Matter,''
Phys. Rev. D \textbf{94}, no.8, 083504 (2016)
[arXiv:1607.06077 [astro-ph.CO]].

\bibitem{Primack:2009jr}
J.~Primack,
``Cosmology: small scale issues revisited,''
New J. Phys. \textbf{11}, 105029 (2009)
[arXiv:0909.2247 [astro-ph.CO]].

\bibitem{Bull:2015stt}
P.~Bull, Y.~Akrami, J.~Adamek, T.~Baker, E.~Bellini, J.~Beltran Jimenez, E.~Bentivegna, S.~Camera, S.~Clesse and J.~H.~Davis, \textit{et al.}
``Beyond $\Lambda$CDM: Problems, solutions, and the road ahead,''
Phys. Dark Univ. \textbf{12}, 56-99 (2016)
[arXiv:1512.05356 [astro-ph.CO]].

\bibitem{Hu:2000ke}
W.~Hu, R.~Barkana and A.~Gruzinov,
``Cold and fuzzy dark matter,''
Phys. Rev. Lett. \textbf{85}, 1158-1161 (2000)
[arXiv:astro-ph/0003365 [astro-ph]].

\bibitem{Khmelnitsky:2013lxt}
A.~Khmelnitsky and V.~Rubakov,
``Pulsar timing signal from ultralight scalar dark matter,''
JCAP \textbf{02}, 019 (2014)
[arXiv:1309.5888 [astro-ph.CO]].

\bibitem{LISA:2017pwj}
P.~Amaro-Seoane \textit{et al.} [LISA],
``Laser Interferometer Space Antenna,''
[arXiv:1702.00786 [astro-ph.IM]].

\bibitem{Aoki:2016kwl}
A.~Aoki and J.~Soda,
``Detecting ultralight axion dark matter wind with laser interferometers,''
Int. J. Mod. Phys. D \textbf{26}, no.07, 1750063 (2016)
[arXiv:1608.05933 [astro-ph.CO]].

\bibitem{Yu:2023iog}
J.~C.~Yu, Y.~H.~Yao, Y.~Tang and Y.~L.~Wu,
``Sensitivity of space-based gravitational-wave interferometers to ultralight bosonic fields and dark matter,''
Phys. Rev. D \textbf{108}, no.8, 8 (2023)
[arXiv:2307.09197 [gr-qc]].

\bibitem{Wang:2023phr}
K.~Wang and Y.~Zhong,
``Frequency modulation of gravitational waves by ultralight scalar dark matter,''
Phys. Rev. D \textbf{108}, no.12, 123531 (2023)
[arXiv:2306.10732 [astro-ph.CO]].

\bibitem{Miller:2025yyx}
A.~L.~Miller,
``Gravitational wave probes of particle dark matter: a review,''
[arXiv:2503.02607 [astro-ph.HE]].

\bibitem{Blas:2016ddr}
D.~Blas, D.~L.~Nacir and S.~Sibiryakov,
``Ultralight Dark Matter Resonates with Binary Pulsars,''
Phys. Rev. Lett. \textbf{118}, no.26, 261102 (2017)
[arXiv:1612.06789 [hep-ph]].

\bibitem{Boskovic:2018rub}
M.~Bo\v{s}kovi\'c, F.~Duque, M.~C.~Ferreira, F.~S.~Miguel and V.~Cardoso,
``Motion in time-periodic backgrounds with applications to ultralight dark matter haloes at galactic centers,''
Phys. Rev. D \textbf{98}, 024037 (2018)
[arXiv:1806.07331 [gr-qc]].

\bibitem{Brito:2020lup}
R.~Brito, S.~Grillo and P.~Pani,
``Black Hole Superradiant Instability from Ultralight Spin-2 Fields,''
Phys. Rev. Lett. \textbf{124}, no.21, 211101 (2020)
[arXiv:2002.04055 [gr-qc]].

\bibitem{Dalal:2022rmp}
N.~Dalal and A.~Kravtsov,
``Excluding fuzzy dark matter with sizes and stellar kinematics of ultrafaint dwarf galaxies,''
Phys. Rev. D \textbf{106}, no.6, 063517 (2022)
[arXiv:2203.05750 [astro-ph.CO]].

\bibitem{Amin:2022nlh}
M.~A.~Amin and M.~Mirbabayi,
``A Lower Bound on Dark Matter Mass,''
Phys. Rev. Lett. \textbf{132}, no.22, 221004 (2024)
[arXiv:2211.09775 [hep-ph]].

\bibitem{Jain:2021pnk}
M.~Jain and M.~A.~Amin,
``Polarized solitons in higher-spin wave dark matter,''
Phys. Rev. D \textbf{105} (2022) no.5, 056019
[arXiv:2109.04892 [hep-th]].

\bibitem{Schiappacasse:2025mao}
E.~D.~Schiappacasse,
``Dark spin-2 field solitons as a source of electromagnetic radiation,''
JCAP \textbf{08} (2025), 085
[arXiv:2503.12569 [hep-ph]].

\bibitem{Hui:2021tkt}
L.~Hui,
``Wave Dark Matter,''
Ann. Rev. Astron. Astrophys. \textbf{59} (2021), 247-289
[arXiv:2101.11735 [astro-ph.CO]].

\bibitem{NANOGrav:2023gor}
G.~Agazie \textit{et al.} [NANOGrav],
``The NANOGrav 15 yr Data Set: Evidence for a Gravitational-wave Background,''
Astrophys. J. Lett. \textbf{951}, no.1, L8 (2023)
[arXiv:2306.16213 [astro-ph.HE]].

\bibitem{EPTA:2023fyk}
J.~Antoniadis \textit{et al.} [EPTA and InPTA:],
``The second data release from the European Pulsar Timing Array - III. Search for gravitational wave signals,''
Astron. Astrophys. \textbf{678}, A50 (2023)
[arXiv:2306.16214 [astro-ph.HE]].

\bibitem{Navarro:1995iw}
J.~F.~Navarro, C.~S.~Frenk and S.~D.~M.~White,
``The Structure of cold dark matter halos,''
Astrophys. J. \textbf{462}, 563-575 (1996)
[arXiv:astro-ph/9508025 [astro-ph]].

\bibitem{Gao:2021sxw}
Z.~Gao, X.~Chen, Y.~M.~Hu, J.~D.~Zhang and S.~J.~Huang,
``A higher probability of detecting lensed supermassive black hole binaries by LISA,''
Mon. Not. Roy. Astron. Soc. \textbf{512}, no.1, 1-10 (2022)
[arXiv:2102.10295 [astro-ph.CO]].

\bibitem{Choi:2021bkx}
H.~G.~Choi, C.~Park and S.~Jung,
``Small-scale shear: Peeling off diffuse subhalos with gravitational waves,''
Phys. Rev. D \textbf{104}, no.6, 063001 (2021)
[arXiv:2103.08618 [astro-ph.CO]].

\bibitem{Cremonese:2021ahz}
P.~Cremonese, D.~F.~Mota and V.~Salzano,
``Characteristic Features of Gravitational Wave Lensing as Probe of Lens Mass Model,''
Annalen Phys. \textbf{535}, no.6, 2300040 (2023)
[arXiv:2111.01163 [astro-ph.CO]].

\bibitem{Guo:2022dre}
X.~Guo and Y.~Lu,
``Probing the nature of dark matter via gravitational waves lensed by small dark matter halos,''
Phys. Rev. D \textbf{106}, no.2, 023018 (2022)
[arXiv:2207.00325 [astro-ph.CO]].

\bibitem{Tambalo:2022wlm}
G.~Tambalo, M.~Zumalac{\'a}rregui, L.~Dai and M.~H.~Y.~Cheung,
``Gravitational wave lensing as a probe of halo properties and dark matter,''
Phys. Rev. D \textbf{108}, no.10, 103529 (2023)
[arXiv:2212.11960 [astro-ph.CO]].

\bibitem{Takahashi:2003ix}
R.~Takahashi and T.~Nakamura,
``Wave effects in gravitational lensing of gravitational waves from chirping binaries,''
Astrophys. J. \textbf{595}, 1039-1051 (2003)
[arXiv:astro-ph/0305055 [astro-ph]].

\bibitem{He:2021hhl}
J.~H.~He,
``gwsim: a code to simulate gravitational waves propagating in a potential well,''
Mon. Not. Roy. Astron. Soc. \textbf{506}, no.4, 5278-5293 (2021)
[arXiv:2107.09800 [gr-qc]].

\bibitem{Qiu:2022dya}
Y.~Qiu, K.~Wang and J.~h.~He,
``Amplitude modulation in binary gravitational lensing of gravitational waves,''
[arXiv:2205.01682 [gr-qc]].

\bibitem{He:2022sjf}
J.~h.~He and Z.~Wu,
``Simulating gravitational waves passing through the spacetime of a black hole,''
Phys. Rev. D \textbf{106}, no.12, 124037 (2022)
[arXiv:2208.01621 [gr-qc]].

\bibitem{Yin:2023kzr}
C.~Yin and J.~h.~He,
``Detectability of Single Spinless Stellar-Mass Black Holes through Gravitational Lensing of Gravitational Waves with Advanced LIGO,''
Phys. Rev. Lett. \textbf{132}, no.1, 011401 (2024)
[arXiv:2312.12451 [astro-ph.HE]].

\bibitem{dealii}
P.~Africa, D.~Arndt and W.~Bangerth, \textit{et al.}
``The deal.II library, Version 9.6,'' 
Journal of Numerical Mathematics \textbf{32}, no.4, 369-380 (2024) 

\bibitem{Guzman:2004wj}
F.~S.~Guzman and L.~A.~Urena-Lopez,
``Evolution of the Schrodinger-Newton system for a selfgravitating scalar field,''
Phys. Rev. D \textbf{69}, 124033 (2004)
[arXiv:gr-qc/0404014 [gr-qc]].

\Mov{\bibitem{Davies:2019wgi}
E.~Y.~Davies and P.~Mocz,
``Fuzzy Dark Matter Soliton Cores around Supermassive Black Holes,''
Mon. Not. Roy. Astron. Soc. \textbf{492} (2020) no.4, 5721-5729
[arXiv:1908.04790 [astro-ph.GA]].

\bibitem{Tan:2024dne}
C.~Tan, M.~L.~Delliou and K.~Wang,
``Diversity of fuzzy dark matter solitons,''
Phys. Rev. D \textbf{112} (2025) no.6, 063021
[arXiv:2411.16114 [hep-ph]].}

\bibitem{Schive:2014hza}
H.~Y.~Schive, M.~H.~Liao, T.~P.~Woo, S.~K.~Wong, T.~Chiueh, T.~Broadhurst and W.~Y.~P.~Hwang,
``Understanding the Core-Halo Relation of Quantum Wave Dark Matter from 3D Simulations,''
Phys. Rev. Lett. \textbf{113}, no.26, 261302 (2014)
[arXiv:1407.7762 [astro-ph.GA]].

\bibitem{Chan:2021bja}
H.~Y.~J.~Chan, E.~G.~M.~Ferreira, S.~May, K.~Hayashi and M.~Chiba,
``The diversity of core\textendash{}halo structure in the fuzzy dark matter model,''
Mon. Not. Roy. Astron. Soc. \textbf{511}, no.1, 943-952 (2022)
[arXiv:2110.11882 [astro-ph.CO]].

\bibitem{Wang:2019ubx}
W.~Wang, J.~Han, M.~Cautun, Z.~Li and M.~N.~Ishigaki,
``The mass of our Milky Way,''
Sci. China Phys. Mech. Astron. \textbf{63}, no.10, 109801 (2020)
[arXiv:1912.02599 [astro-ph.GA]].

\bibitem{Peters:1974gj}
P.~C.~Peters,
``Index of refraction for scalar, electromagnetic, and gravitational waves in weak gravitational fields,''
Phys. Rev. D \textbf{9}, 2207-2218 (1974)
doi:10.1103/PhysRevD.9.2207

\bibitem{Baraldo:1999ny}
C.~Baraldo, A.~Hosoya and T.~T.~Nakamura,
``Gravitationally induced interference of gravitational waves by a rotating massive object,''
Phys. Rev. D \textbf{59}, 083001 (1999)

\Mov{
\bibitem{NANOGrav:2023hde}
G.~Agazie \textit{et al.} [NANOGrav],
``The NANOGrav 15 yr Data Set: Observations and Timing of 68 Millisecond Pulsars,''
Astrophys. J. Lett. \textbf{951} (2023) no.1, L9
[arXiv:2306.16217 [astro-ph.HE]].

\bibitem{Nesti:2013uwa}
F.~Nesti and P.~Salucci,
``The Dark Matter halo of the  Milky Way, AD 2013,''
JCAP \textbf{07} (2013), 016
[arXiv:1304.5127 [astro-ph.GA]].
}
\bibitem{NANOGrav:2023tcn}
G.~Agazie \textit{et al.} [NANOGrav],
``The NANOGrav 15 yr Data Set: Search for Anisotropy in the Gravitational-wave Background,''
Astrophys. J. Lett. \textbf{956}, no.1, L3 (2023)
[arXiv:2306.16221 [astro-ph.HE]].

\end{thebibliography}
\end{document}